\documentstyle[preprint,aps]{revtex}
\begin{document}
\preprint{\begin{minipage}{2in}\begin{flushright}
  Edinburgh Preprint: 94/1 \\[-3mm] hep-lat/9406002 \end{flushright}
  \end{minipage}}
\input epsf
\draft
\title{Radiative corrections to the kinetic couplings
\\in nonrelativistic lattice QCD}
\author{Colin J. Morningstar}
\address{Department of Physics \& Astronomy, University of Edinburgh,
  Edinburgh EH9 3JZ, Scotland}
\date{May 27, 1994}
\maketitle
\begin{abstract}
The heavy-quark mass and wave function renormalizations, energy shift,
and radiative corrections to two important couplings, the so-called
kinetic couplings, in nonrelativistic lattice QCD are determined
to leading order in tadpole-improved perturbation theory.  The scales
at which to evaluate the running QCD coupling for these quantities,
except the wave function renormalization, are obtained using the
Lepage-Mackenzie prescription.  When the bare quark mass is greater
than the inverse lattice spacing, the kinetic coupling corrections
are roughly 10\% of the tree-level coupling strengths; these corrections
grow quickly as the bare quark mass becomes small.  A need for computing
the two-loop corrections to the energy shift and mass renormalization
is demonstrated.
\end{abstract}

\narrowtext
\section{Introduction}
\label{sec:intro}

Nonrelativistic lattice QCD (NRQCD) is an effective field theory
formulated to reproduce the action of continuum QCD at low energies
\cite{lep91,lep92}.  The NRQCD Lagrangian includes interactions
which systematically correct for relativity and finite-lattice-spacing
errors.  To complete the formulation of NRQCD, the coupling strengths
of its interactions must be determined.  One
possibility is to treat the couplings as adjustable
parameters and tune them to fit certain experimental data;
however, this tuning significantly reduces the predictive power of
NRQCD simulations and is very costly and difficult.  A better
alternative is to compute the couplings in terms of the fundamental
QCD coupling $\alpha_s=g^2/(4\pi)$ and the bare heavy-quark mass $M$
using perturbation theory.  This is done by evaluating various
scattering amplitudes in both QCD and lattice NRQCD and adjusting
the couplings until these amplitudes agree at low energies.
Since the role of these couplings is to compensate for neglected
low-energy effects from highly-ultraviolet QCD processes, one expects
that they may be computed to a good approximation using perturbation
theory, provided that $a\Lambda_{\rm QCD}$ is small, where $a$ is
the lattice spacing and $\Lambda_{\rm QCD}$ is the QCD scale parameter.

In this paper, the heavy-quark propagator in NRQCD is calculated
to leading order in tadpole-improved perturbation theory and the
on-mass-shell dispersion relation for the heavy quark is obtained.
The mass renormalization, a shift in the zero point of energy, and
the radiative corrections to the couplings of two important NRQCD
interactions are determined by matching this dispersion relation to
that of continuum QCD to $O(v^4)$, where $v$ is the expectation value
of the heavy-quark velocity in a typical heavy-quark hadron.
The wave function renormalization is extracted from the on-shell
residue of the perturbative propagator. These parameters are needed
for high-precision numerical simulations of quarkonium and heavy-light
mesons.  Results are obtained using the standard Wilson action and a
tree-level $O(a^2)$-improved action for the gluons, and using two
versions of the NRQCD action: one containing all $O(v^2)$ spin-independent
and all $O(v^4)$ spin-dependent corrections to the leading kinetic term
with tree-level removal of cutoff errors, the other including only
$O(v^2)$ corrections with no tree-level removal of lattice-spacing
errors.

In addition, the scales at which to evaluate the running
coupling for these quantities, except the wave function renormalization,
are obtained for the first time using
the Lepage-Mackenzie prescription.  Choosing a value for
$a\Lambda_{\rm QCD}$ from recent simulation measurements of the QCD
coupling, numerical estimates of the energy shift, mass renormalization,
and the radiative corrections to the two so-called kinetic couplings
are also obtained.  When $aM>1$, where $M$ is the bare quark mass, the
radiative corrections to the kinetic couplings are roughly 10\% of
the tree-level coupling strengths; these corrections grow quickly as $aM$
becomes small.  A range of estimates for the energy shift and mass
renormalization are obtained after circumventing certain defects in the
Lepage-Mackenzie prescription for setting the scale.  The mass
renormalization is small for $aM>3$ and becomes large as $aM$ decreases
below this.
Problems in reliably setting the scale result in large uncertainties in
determining the energy shift and mass renormalization for small $aM$.
These uncertainties underscore the need to compute the two-loop
corrections to these renormalization parameters.

This work is partly
an extension of previous calculations \cite{dav92,mor93} and is being
carried out in conjunction with ongoing simulations as part of the
NRQCD collaboration \cite{nrqcdcollab,mbpaper}.

The heavy-quark propagator in lattice NRQCD is briefly
described in Sec.~\ref{sec:nrqcd}.
The energy shift, mass and wave function renormalizations, and the kinetic
coupling corrections are defined and their calculation is outlined in
Sec.~\ref{sec:couplings}.  The determination of the Lepage-Mackenzie
scales is also described in this section.  Results of these
calculations are presented in Sec.~\ref{sec:results}.  A final summary
and suggestions for future work are given in Sec.~\ref{sec:concl}.
\section{Quark propagator in lattice NRQCD}
\label{sec:nrqcd}

The heavy-quark propagator ${\cal G}(x)=a^3\langle \psi(x)
\psi^\dagger(0)\rangle$ in lattice NRQCD may be defined such that it
satisfies an evolution equation given by \cite{change}
\begin{eqnarray}
\label{evoleqn}
{\cal G}(x\!+\!a{\hat e}_4) &=& u_0^{-1}U^\dagger_4(x)\
[O(x){\cal G}(x)+\delta(x;0)], \nonumber\\
O(x) &=& \Bigl(1\!-\!{aH_0 \over 2n}\Bigr)^n\!
(1\!-\!a \delta H)\Bigl(1\!-\!{aH_0 \over 2n}\Bigr)^n,\\
{\cal G}(x) &=& 0, \quad\quad (x_4\le 0)\nonumber
\end{eqnarray}
where $a$ is the lattice spacing, $n$ is a positive integer\cite{lep91},
$\psi(x)$ is the heavy-quark field, $U_\mu(x)$ is the link variable
representing the gauge field along the link between sites $x$
and $x+a{\hat e}_\mu$
with $({\hat e}_\mu)_\alpha=\delta_{\mu\alpha}$,
and $u_0$ is a mean-field parameter \cite{lep92b}
defined in terms of the average plaquette.
The nonrelativistic kinetic energy operator $H_0$ is given by
\begin{equation}
H_0 = -{{\bf\Delta}^{(2)} \over 2M},
\end{equation}
and the relativistic and finite-lattice-spacing corrections are
included with coupling strengths $c_j$ in
\begin{equation}
\delta H = \sum_{j=1}^8 c_j\ V_j,
\label{dHeqn}
\end{equation}
where
\begin{eqnarray}
V_1 &=& - {({\bf\Delta}^{(2)})^2\over 8M^3}\Bigl(1+\frac{Ma}{2n}
        \Bigr), \\
V_2 &=& \frac{a^2{\bf\Delta}^{(4)}}{24M},\\
V_3 &=& {ig\over 8M^2} \bigl({\bf\Delta}^{(\pm)}\cdot {\bf E}
      - {\bf E} \cdot {\bf\Delta}^{(\pm)}  \bigl), \\
V_4 &=& - {g\over 8M^2} {\hbox{\boldmath $\sigma$}}\cdot{\bf(\tilde
        \Delta^{(\pm)}\times \tilde E - \tilde E \times
        \tilde\Delta^{(\pm)})}, \label{V4eqn} \\
V_5 &=& - {g\over 2M} {\hbox{\boldmath $\sigma$}}\cdot{\bf\tilde B},
        \label{V5eqn} \\
V_6 &=& - {g \over 8M^3}\lbrace{\bf\Delta}^{(2)},
        {\hbox{\boldmath $\sigma$}}
  \cdot {\bf B}\rbrace, \\
V_7 &=& - {3g\over 64M^4}\lbrace{\bf\Delta}^{(2)},
        {\hbox{\boldmath $\sigma$}}
         \cdot({\bf\Delta^{(\pm)}\!\times\! E\! -\!
         E\!\times\!\Delta^{(\pm)})}
         \rbrace, \\
V_8 &=& - {ig^2\over 8M^3} {\hbox{\boldmath $\sigma$}}
        \cdot{\bf E\times E}.
\label{Vjeqn}
\end{eqnarray}
The covariant, tadpole-improved difference operators are defined by
\begin{eqnarray}
a\Delta_\mu^{(+)} &=& u_0^{-1}U_\mu - 1,\\
a\Delta_\mu^{(-)} &=& 1-u_0^{-1} U^\dagger_\mu,\\
\Delta^{(\pm)} &=& \frac{1}{2}(\Delta^{(+)}+\Delta^{(-)}),\\
a^{2m} {\bf\Delta}^{(2m)} &=& \sum_{k=1}^3
\left( u_0^{-1} [U_k + U^\dagger_k] - 2 \right)^m,
\end{eqnarray}
and $\bf E$ and $\bf B$ are the cloverleaf, mean-field-improved
chromoelectric and
chromomagnetic fields, respectively\cite{man83}.
Tree-level removal of the leading cutoff errors in these operators
\cite{lep92} produces the operators $\bf\tilde\Delta^{(\pm)}$,
${\bf\tilde\Delta}^{(2)}$, $\bf\tilde E$, and $\bf\tilde B$.
The components of ${\hbox{\boldmath $\sigma$}}$ are the standard
Pauli spin matrices. The parameter $n$ eliminates doublers
and stabilizes the evolution of the quark Green's function
when $n\gtrsim 3/(Ma)$.

The coupling coefficients $c_j$ are functions of $g$ and $aM$
and are determined by matching low-energy scattering amplitudes
in lattice NRQCD with those from continuum QCD, order-by-order in
perturbation theory.  At tree level, the values of these couplings
are unity\cite{lep92}.  The $O(\alpha_s)$ matching determines the
radiative corrections to the $c_j$ couplings and may possibly
require the introduction of additional $O(a^2)$-suppressed
interactions.

The operators $V_1$ and $V_2$, although appropriately gauged, will
be referred to here as {\em kinetic
operators} because they affect the energy of the free quark in the
absence of the gluon fields.  Their associated couplings $c_1$ and
$c_2$ will be called {\em kinetic couplings}.

\section{Kinetic couplings and renormalizations}
\label{sec:couplings}

The kinetic couplings $c_1$ and $c_2$, the energy shift, and the
heavy-quark mass and wave function renormalizations may be
determined by studying the heavy-quark propagator in perturbation
theory.  An important consideration in the perturbative study of this
propagator is the choice of expansion parameter.  Lepage and
Mackenzie have advocated a renormalized running QCD coupling
defined in terms of the short-distance static-quark potential
and evaluated at a prescribed mass scale.  Their coupling shall
be adopted here.

In this section, Eq.~\ref{evoleqn} is solved perturbatively to
$O(\alpha_s)$.  The dispersion relation obtained from this solution
is then matched to that of continuum QCD to $O(v^4)$ to determine
the heavy-quark mass renormalization, the shift in the zero point
of energy, and the radiative corrections to $c_1$ and $c_2$.  The
wave function renormalization is then obtained from the on-shell
residue of the perturbative propagator.  Lastly, a brief description
of the Lepage-Mackenzie coupling $\alpha_V(q^\ast)$ is presented
and their procedure of setting the scale $q^\ast$ is applied to
the renormalization parameters and kinetic couplings.

\subsection{Heavy-quark self-energy}
\label{sec:calc}
\begin{figure}
\begin{center}
\leavevmode
\epsfbox[80 300 530 760]{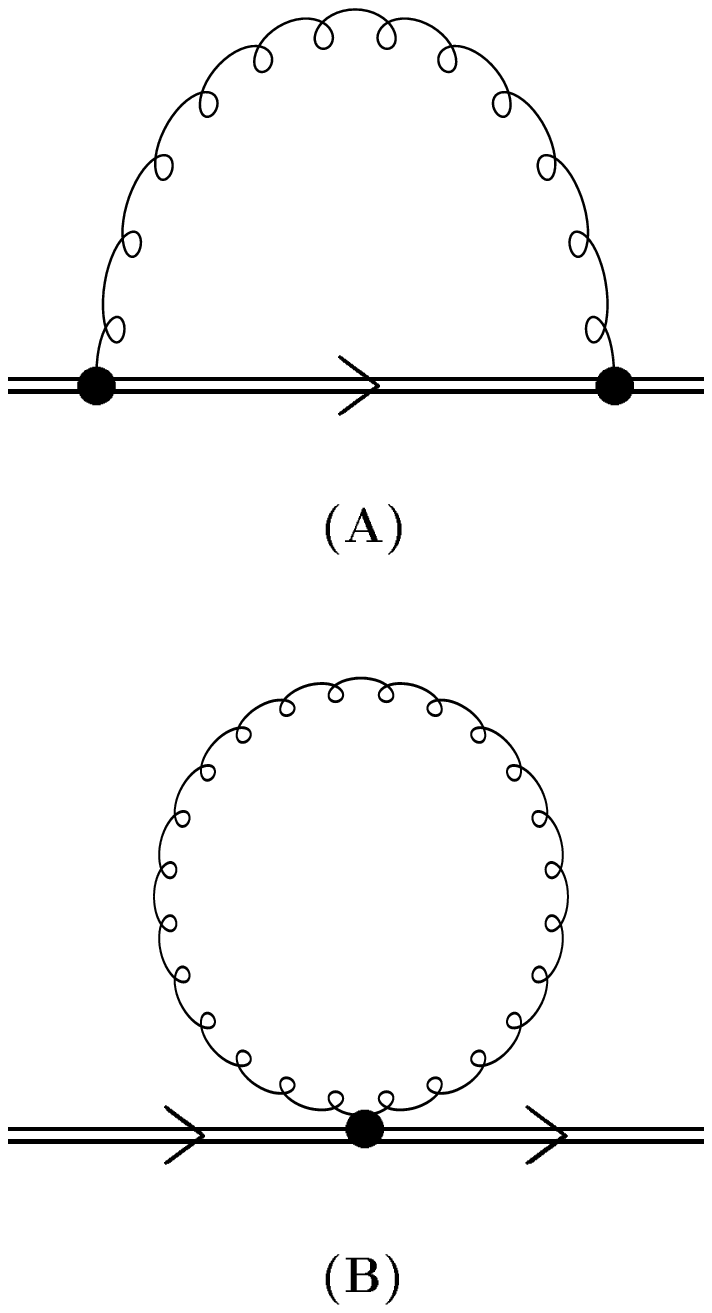}
\end{center}
\caption[figdiag]{Two Feynman diagrams which
contribute to the heavy-quark self-energy.  A curly line denotes
a gluon; a double solid line denotes a heavy quark.}
\label{figdiag}
\end{figure}

The heavy-quark self-energy $\Sigma(p)$ is defined by writing the
full inverse quark propagator ${\cal G}^{-1}(p)$ in the form
\begin{equation}
a {\cal G}^{-1}(p)_{\alpha\beta}^{ij} =
Q^{-1}(p)\delta^{ij}\delta_{\alpha\beta}
 - a \Sigma^{ij}_{\alpha\beta}(p),
\end{equation}
where $i,j$ are color indices, $\alpha,\beta$ are spin indices,
$p=({\bf p},p_4)$ is a four-momentum in Euclidean space, and $Q(p)$ is
the zeroth-order heavy-quark propagator in momentum space given by:
\begin{eqnarray}
 Q(p)&=& \Bigl[
 e^{ip_4a}- \Lambda_n({\bf p})\ \Gamma_n{(\bf p})^{2n} \Bigr]^{-1}, \\
 \Gamma_n({\bf p}) &=& 1-\frac{\kappa_2({\bf p})}{nMa},\\
 \Lambda_n({\bf p}) &=&
1+{2\kappa_2({\bf p})^2\over M^3a^3}\Bigl(1+\frac{Ma}{2n}\Bigr)
-{2\kappa_4({\bf p})\over 3Ma},
\end{eqnarray}
with $\kappa_n({\bf p}) = \sum_{j=1}^3 \sin^n\left(p_ja/2\right)$.
At leading order in perturbation theory, this self-energy is given by:
\begin{equation}
\Sigma^{ij}_{\alpha\beta}(p)=
\alpha_s\delta^{ij} \delta_{\alpha\beta}
\biggl[\Sigma^{(A)}(p)\!+\!\Sigma^{(B)}(p)
\!+\!\Sigma^{(C)}(p)\biggr],
\end{equation}
where $\Sigma^{(A)}(p)$ denotes the contribution from
the quark-gluon loop diagram shown in Fig.~\ref{figdiag}(A),
$\Sigma^{(B)}(p)$ represents the contribution from the
tadpole graph shown in Fig.~\ref{figdiag}(B), and $\Sigma^{(C)}(p)$
denotes the sum of the order $\alpha_s$ contributions from the $c_1$,
$c_2$, and link-variable renormalization counterterms which
arise when one writes $u_0=1+\alpha_s u_0^{(2)}+O(\alpha_s^2)$ and
$c_j=1+\alpha_s c_j^{(2)}+O(\alpha_s^2)$.

The self-energy $\Sigma(p)$ is invariant under
interchange of any two spatial momentum components
$p_i\leftrightarrow p_j$ and under spatial momentum reflections
$p_j \rightarrow -p_j$, and transforms into its complex conjugate
under $p_4\rightarrow -p_4^\ast$.  From these properties, the
small-${\bf p}$ representation of the self-energy may be written:
\begin{eqnarray}
\label{selfenexp}
  a\Sigma(p) &\approx& \alpha_s\Bigl\lbrace f_0(w) + f_1(w)
{{\bf p}^2a^2\over 2Ma} + f_2(w)\frac{({\bf p}^2)^2a^4}{8M^2a^2}
\nonumber\\ &+& f_3(w)\ (\sum_{j=1}^3p_j^4a^4) +\dots\Bigr\rbrace,
\end{eqnarray}
where $w=-ip_4a$.  When $w$ is small, the functions $f_m(w)$ are
represented well by their power series expansions
$f_m(w) = \sum_{l=0}^\infty \Omega_m^{(l)}\ w^l$.

The moments $\Omega_m^{(l)}$ are obtained from integrals of the form
\begin{equation}
\Omega_m^{(l)} = \int_{-\pi/a}^{\pi/a} \frac{d^4q}{(2\pi)^4}\
 \xi_m^{(l)}(q),
\label{intform}
\end{equation}
where $q$ is the momentum of the gluon in the loop diagrams shown
in Fig.~\ref{figdiag} and $\xi_m^{(l)}(q)$ is some combination of
vertex factors and Feynman propagators.
To evaluate these integrals \cite{mor93diff},
first, a small gluon mass $\lambda$ is introduced to provide an
infrared cutoff.  Due to singularities in the integrands from the
zeroth-order quark propagator, it is best to first evaluate the
integrals over the temporal component of $q$ analytically.
Using $z=\exp(-iq_4a)$, the $q_4$ integrations are transformed into
contour integrals along the $\vert z\vert =1$ unit circle which can
be evaluated by the residue theory.
The remaining integrals over the spatial components of $q$ are then
approximated numerically by discrete sums using $aq_j = 2\pi n_j/N$,
where the $n_j$ take all integer values satisfying $-N/2<n_j\le N/2$
for integral $N$.  The error made in this approximation diminishes
exponentially fast as $N\rightarrow \infty$, but the rate of
decay is proportional to $\lambda$.  Convergence in $N$ can be
dramatically improved by making the following change in the
variables: $q_j\rightarrow q_j -\alpha \sin q_j$, where
$\alpha={\rm sech} u$ and $u$ satisfies $a\lambda = u-{\rm tanh} u$.
Extrapolation of the results to zero gluon mass is accomplished using a
least-squares fit to the form: $\sum_{m=0}^N b_m^{(0)} (a\lambda)^m
+\sum_{m=1}^M b_m^{(1)} (a\lambda)^m \ln a^2\lambda^2$.
Calculations are performed in the Feynman gauge.

The contributions from $\Sigma^C(p)$ to the $f_m(w)$ functions
appearing in Eq.~\ref{selfenexp} can be exactly determined.
One finds
\begin{eqnarray}
 f_0^{(C)}(w) &=& -u_0^{(2)}\biggl(e^{-w}\!+\!\frac{3}{Ma}\biggr),\\
 f_1^{(C)}(w) &=& \frac{2}{3}u_0^{(2)}\biggl(1\!+\!\frac{9}{2Ma}
   \!+\!\frac{9}{2M^2a^2}\biggr),\\
 f_2^{(C)}(w) &=& -\frac{4}{3}u_0^{(2)}\biggl(1\!+\!\frac{3}{8Ma}
   \Bigl(10\!+\!\frac{3}{n^2}\Bigr)\!+\!
   \frac{27}{4M^2a^2}\biggr)\nonumber\\
        &+& c_1^{(2)} \biggl(\frac{1}{2n}\!
       +\!\frac{1}{Ma}\biggr),\\
 f_3^{(C)}(w) &=& \frac{u_0^{(2)}}{18Ma}
   \biggl(1\!-\!\frac{9}{4M^2a^2}\biggr)
   \!-\!\frac{c_2^{(2)}}{24Ma}.
\end{eqnarray}
Contributions to the moments $\Omega_m^{(l)}$ from the
$c_1$, $c_2$, and mean-field improvement counterterms in
$\Sigma^{(C)}(p)$ are then easily determined from the power
series expansions of these functions.

\subsection{Matching}
{}From the heavy-quark propagator computed to $O(\alpha_s)$ in
perturbation theory, one finds that the on-mass-shell quark satisfies
a dispersion relation given to order $v^4$ by
\begin{equation}
 \omega_0({\bf p}) \approx {{\bf p}^2\over 2M_r}
 - {({\bf p}^2)^2\over 8M_r^3} - \alpha_s
 \delta\omega({\bf p})+\dots,
\label{dispers}
\end{equation}
where $\omega_0({\bf p})$ is the value of $w=-ip_4a$ at which the
inverse propagator vanishes, $M_r$ is the renormalized mass given by
$M_r=Z_m M$ with $Z_m=1+\alpha_s(\Omega_0^{(0)}+\Omega_0^{(1)}
+\Omega_1^{(0)})$, and
\begin{eqnarray}
a\delta\omega({\bf p}) &=& W_0 + W_1\ \frac{({\bf p}^2)^2a^4}{8M^2a^2}
 + W_2\ (\sum_{j=1}^3p_j^4a^4),\\
W_0 &=& \Omega_0^{(0)},\\
W_1 &=& \Omega_0^{(0)}\left(1+\frac{2}{Ma}\right) +2 \Omega_0^{(1)}
  \left(1+\frac{1}{Ma}\right)+ 2\Omega_0^{(2)} \nonumber\\
  &+& \Omega_1^{(0)}\left(2+\frac{3}{Ma}\right) + 2\Omega_1^{(1)}
  +\Omega_2^{(0)},\\
W_2 &=& \Omega_3^{(0)}.
\end{eqnarray}
Note that the rest mass $M_r$ has been removed from the
heavy-quark energy in Eq.~\ref{dispers}.

The order $\alpha_s$ corrections to the heavy-quark propagator in
QCD only renormalize the quark field and mass to order $v^4$ so that
the dispersion relation for the quark in continuum QCD is given by
\begin{equation}
 \omega_{\rm QCD}({\bf p}) \approx {{\bf p}^2\over 2M_r}
 - {({\bf p}^2)^2\over 8M_r^3} +\dots\ .
\end{equation}
If lattice NRQCD is to reproduce the low-energy physical predictions of
full QCD, then $\delta\omega({\bf p})=0$ is required in Eq.~\ref{dispers};
that is, $W_0=W_1=W_2=0$.  Alternatively,
since the constant term $W_0$ represents an overall energy
shift, one could require only $W_1=W_2=0$
and then simply shift the energies obtained in simulations by
an amount $\alpha_s W_0/a+M_r$ for each heavy quark.
Setting $W_1=W_2=0$ determines the
coefficients $c_1^{(2)}$ and $c_2^{(2)}$, and defining
$\bar p_4=p_4+i\alpha_s W_0/a$, the inverse propagator
for small $v$ may then be written:
\begin{equation}
 a{\cal G}^{-1}(p)\approx Z_\psi^{-1} \left(
 i \bar p_4a + {{\bf p}^2a^2\over 2M_ra} + \dots\right),
\end{equation}
where $Z_\psi=1-\alpha_s(\Omega_0^{(0)}+\Omega_0^{(1)})
+O(v^2,\alpha_s v^2)$ is the wave function renormalization.

A more convenient set of renormalization parameters may be obtained
by defining
\begin{eqnarray}
 \bar p_4 &=& p_4-i\alpha_s A/a, \\
 Z_m &=& 1+\alpha_s B, \\
 Z_\psi &=& 1+\alpha_s(C+C_{\rm div}),
\end{eqnarray}
where $C_{\rm div}=-2(\ln a^2\lambda^2)/(3\pi)$ is the
infrared-divergent portion of the wave function renormalization.
The parameters $A$, $B$, and $C$ can then be calculated using
$A=-\Omega_0^{(0)}$, $B=\Omega_0^{(0)}+\Omega_0^{(1)}
+\Omega_1^{(0)}$, and $C=-(\Omega_0^{(0)}+\Omega_0^{(1)}+C_{\rm div})$.

\subsection{Expansion parameter $\alpha_V(q^\ast)$}
In order to obtain numerical values for the renormalization
parameters and the kinetic couplings, one must determine the value
of the renormalized QCD coupling $\alpha_s$.  This can be done only
after a definition of the running coupling and a procedure
for determining the relevant mass scale are specified.

A renormalization scheme, due to Lepage and Mackenzie\cite{lep92b},
which defines
the coupling such that the short-distance static-quark potential
has no $\alpha_s^2$ or higher-order corrections is particularly
attractive.  By absorbing higher-order contributions to the
static-quark potential into the coupling, it is hoped that higher-order
contributions to other physical quantities in terms of this coupling
will then be small. In this scheme, the running coupling $\alpha_V(q)$
is defined by
\begin{equation}
V(q)\equiv -\frac{4\pi C_F}{q^2} \alpha_V(q)
\label{coupdef}
\end{equation}
for large $q$, where $C_F=4/3$ is the quark's color Casimir and
$V(q)$ is the static-quark potential at momentum transfer $q$.
For sufficiently large $q$, this coupling runs according to the usual
two-loop relation
\begin{equation}
\alpha_V(q) = \left[\beta_0\ln(q^2/\Lambda_V^2)+
 \frac{\beta_1}{\beta_0}\ln\ln(q^2/\Lambda_V^2) \right]^{-1},
\label{twolooprun}
\end{equation}
where $\beta_0=(11-\frac{2}{3}N_F)/(4\pi)$, $\beta_1=(102-
\frac{38}{3}N_F)/(16\pi^2)$, $\Lambda_V$ is the QCD scale parameter,
and $N_F$ is the number of dynamical quark flavors at mass scale $q$.

Lepage and Mackenzie have also devised a simple procedure for
determining the scale $q^\ast$ at which to evaluate the coupling
$\alpha_V(q^\ast)$ for a given one-loop process \cite{lep92b}.
For a one-loop perturbative contribution of the form
\begin{equation}
I_\xi = \alpha_V(q^\ast) \int \frac{d^4q}{(2\pi)^4}\ \xi(q),
\end{equation}
where $q$ is the momentum of the exchanged gluon, they suggest
\begin{equation}
 \ln(q^{\ast 2})\equiv \frac{\int d^4q \ln(q^2)\ \xi(q)}{\int d^4q\ \xi(q)}.
\label{scaleform}
\end{equation}
Clearly, difficulties with this definition will arise when
$\int d^4q\ \xi(q) \sim 0$.  Also note that the mean value theorem
guarantees that $q^\ast$ will satisfy $0\leq aq^\ast \leq 2\pi$ only
if $\xi(q)\geq 0$ for all $q$ throughout the region of integration
$-\pi \leq aq_\mu \leq \pi$ (or $\xi(q)\leq 0$ for all such $q$).

To evaluate the scales $q^\ast_A$, $q^\ast_B$, $q^\ast_{c_1^{(2)}}$, and
$q^\ast_{c_2^{(2)}}$ corresponding to the gauge-invariant quantities
$A$, $B$, $c_1^{(2)}$, and $c_2^{(2)}$, respectively, integrals of the form
\begin{equation}
\hat\Omega_m^{(l)} = \int \frac{d^4q}{(2\pi)^4}\ \ln(q^2)\ \xi_m^{(l)}(q),
\label{lnintform}
\end{equation}
must be calculated (recall Eq.~\ref{intform}).  First, a small gluon
mass $\lambda$ is introduced to circumvent infrared difficulties.  Next,
the $\ln(a^2q^2)$ factor is approximated by $\ln\{\zeta(z)+a^2({\bf q}^2
+\lambda^2)\}$, where
\begin{equation}
\zeta(z)= -15\ \frac{31z^4+128z^3-318z^2+128z+31}{23z^4+688z^3
           +2358z^2+688z+23},
\label{pade}\end{equation}
and $z=\exp(-iq_4a)$.  Eq.~\ref{pade} is simply a $[2/2]$ Pade
approximant of $a^2q_4^2$ in terms of $\cos(aq_4)$.
This introduces a small error in the determination
of the scale, but enables an analytical treatment of the integration
over the temporal component of $q$.  Singularities in the integrands
make such a treatment crucial to reliably and efficiently evaluating
these integrals.  The remaining integrations over the spatial components
of $q$ are then approximated numerically by discrete sums, as described
previously in Sec.~\ref{sec:calc}, and the $\lambda\rightarrow 0$ limits
are finally taken.

\section{Results and discussion}
\label{sec:results}

Results for $A$, $B$, $C$, $c_1^{(2)}$, $c_2^{(2)}$ and the scales
$q^\ast_A$, $q^\ast_B$, $q^\ast_{c_1^{(2)}}$, $q^\ast_{c_2^{(2)}}$
are presented in this section.  Since the wave function renormalization
is not a gauge-invariant quantity, the scale corresponding to $C$ will
not be considered here.  Choosing a value for $a\Lambda_V$ from recent
simulation measurements of $\alpha_V$, numerical estimates of the
energy shift, mass renormalization, and the radiative corrections
to the kinetic couplings are also obtained.

\subsection{Perturbative coefficients}
Values for $A$, $B$, $C$, $c_1^{(2)}$, and $c_2^{(2)}$ have been
determined for a large range of bare quark masses
and for various values of the stability parameter $n$.
Results were obtained using the standard Wilson action $S_G^{(W)}$
for the gluons and a tree-level $O(a^2)$-improved gluonic action
$S_G^{(I)}$ \cite{lep92}.  In addition, two versions of the
NRQCD interaction operator $\delta H$ were used: one,
denoted by $\delta H^{(4)}$, containing
the full set of interactions $V_1-V_8$ in $\delta H$
(see Eqs.~\ref{dHeqn}$-$\ref{Vjeqn}); the other,
denoted by $\delta H^{(2)}$, including only the $O(v^2)$ corrections
$V_1-V_5$ with no tree-level removal of the cutoff errors
in $V_4$ and $V_5$ (Eqs.~\ref{V4eqn} and \ref{V5eqn} with the
tildes removed).  Contributions to these parameters from the
tadpole-improvement counterterm are given by
\begin{eqnarray}
\delta A &=& u_0^{(2)}\left(1+\frac{3}{Ma}\right),\label{deltaA}\\
\delta B &=& u_0^{(2)}\left(\frac{2}{3}+\frac{3}{M^2a^2}\right),
 \label{deltaB}\\
\delta C &=& u_0^{(2)}\frac{3}{Ma},\label{deltaC}\\
\delta c_1^{(2)} &=& \frac{u_0^{(2)}}{Ma+2n}\left(
 \frac{3}{n}-\frac{18n}{M^2a^2}\right),\label{deltac1}\\
\delta c_2^{(2)} &=& \frac{4}{3}u_0^{(2)}\left(1
 -\frac{9}{4M^2a^2}\right)\label{deltac2}.
\end{eqnarray}
The mean-field correction is $u_0^{(2)}=-\epsilon\pi/3$,
where $\epsilon=1$ for the simple gluonic action and
$\epsilon=0.78463$ for the improved gluonic action.

\begin{figure}
\begin{center}
\leavevmode
\epsfbox[80 360 530 760]{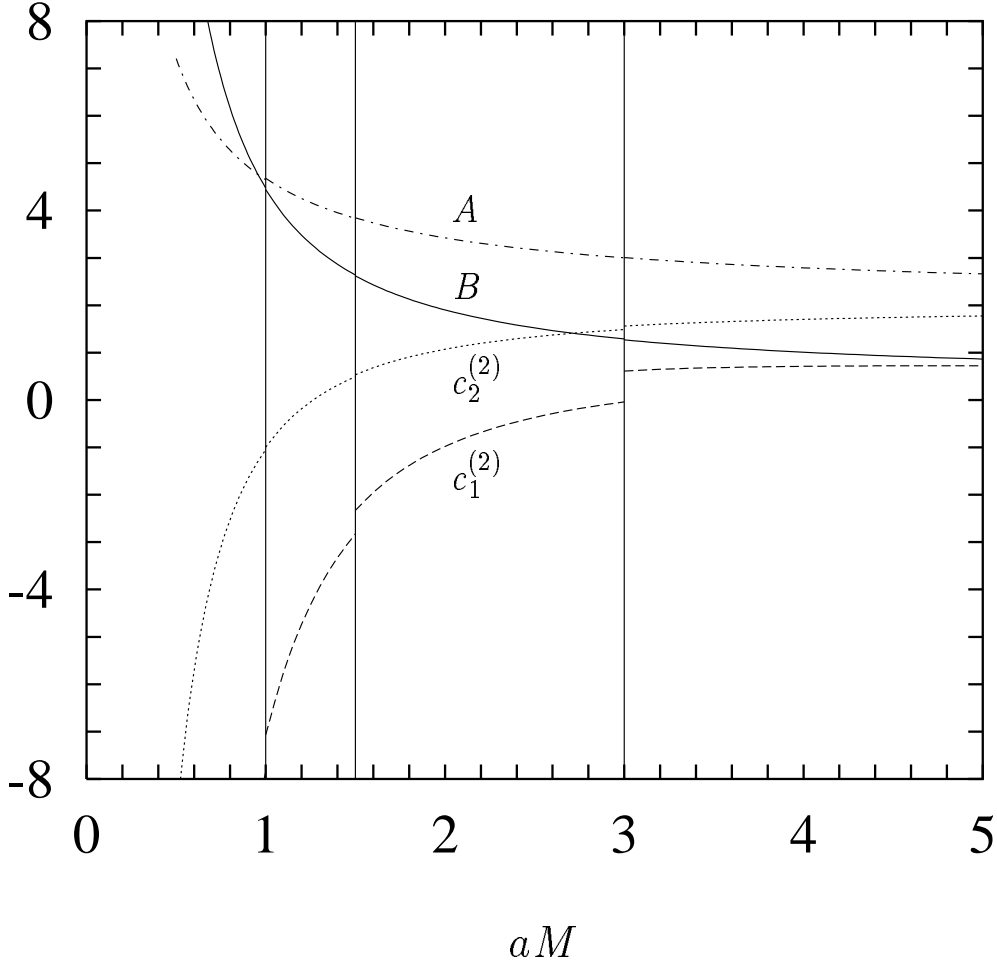}
\end{center}
\caption[figcoefs]
{The energy shift parameter $A$, heavy-quark mass renormalization
parameter $B$, and the kinetic coupling coefficients $c_1^{(2)}$ and
$c_2^{(2)}$ {\em before} tadpole improvement ($u_0=1$) against the
product of the bare heavy-quark mass $M$ and the lattice spacing $a$
using $S_G^{(W)}$ and $\delta H^{(2)}$.  The dot-dashed curve is $A$,
the solid curve shows $B$, the dashed curve is $c_1^{(2)}$,
and the dotted curve indicates $c_2^{(2)}$.
For $aM > 3$, the stability parameter $n$ is set to unity;
for $1.5 < aM < 3$, $n=2$ is used; for $1 < aM < 1.5$, $n=3$ is
used; and for $0.5 < aM < 1$, $n=6$ is used.}
\label{figcoefs}
\end{figure}

\begin{figure}
\begin{center}
\leavevmode
\epsfbox[80 360 530 760]{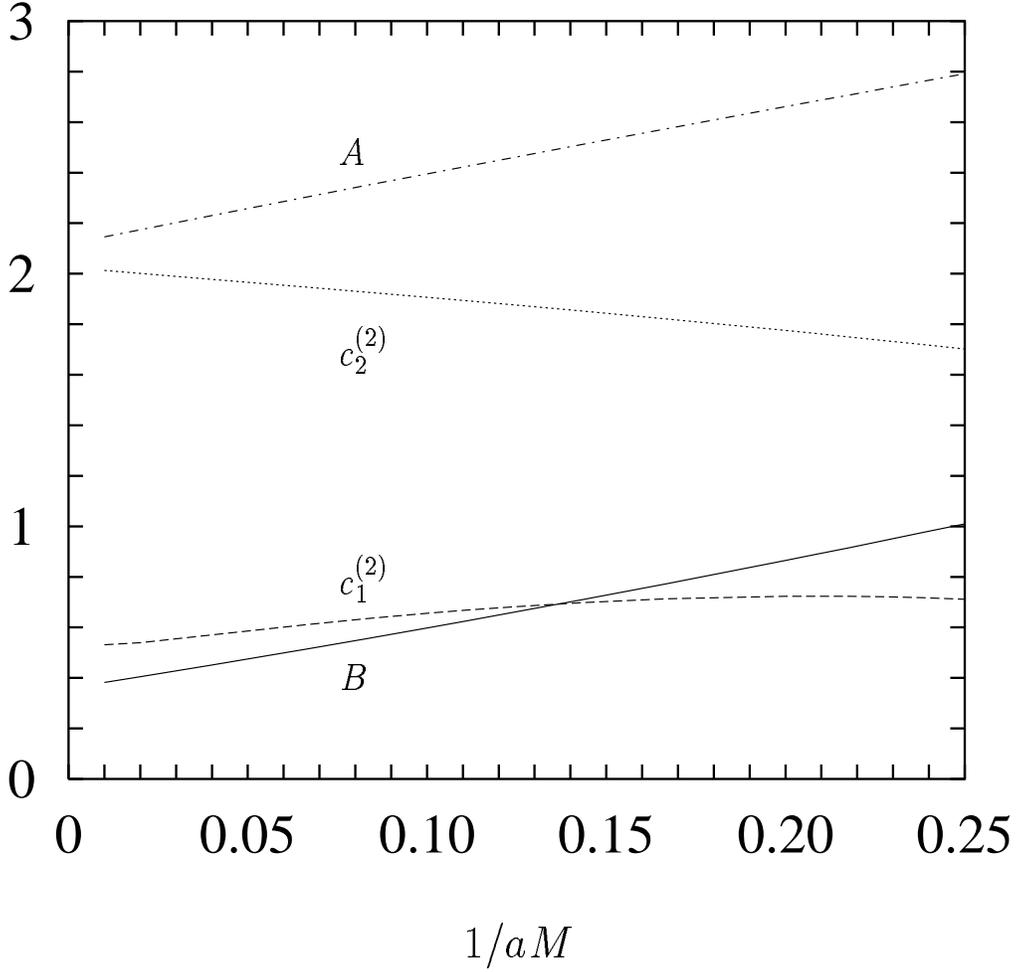}
\end{center}
\caption[figcoefs2]
{The energy shift parameter $A$, heavy-quark mass renormalization
parameter $B$, and the kinetic coupling coefficients $c_1^{(2)}$ and
$c_2^{(2)}$ {\em before} tadpole improvement ($u_0=1$) against $1/aM$
using $S_G^{(W)}$ and $\delta H^{(2)}$.  The dot-dashed curve is $A$,
the solid curve shows $B$, the dashed curve is $c_1^{(2)}$,
and the dotted curve indicates $c_2^{(2)}$.
The stability parameter $n$ is set to unity.}
\label{figcoefs2}
\end{figure}

\begin{figure}
\begin{center}
\leavevmode
\epsfbox[80 360 530 760]{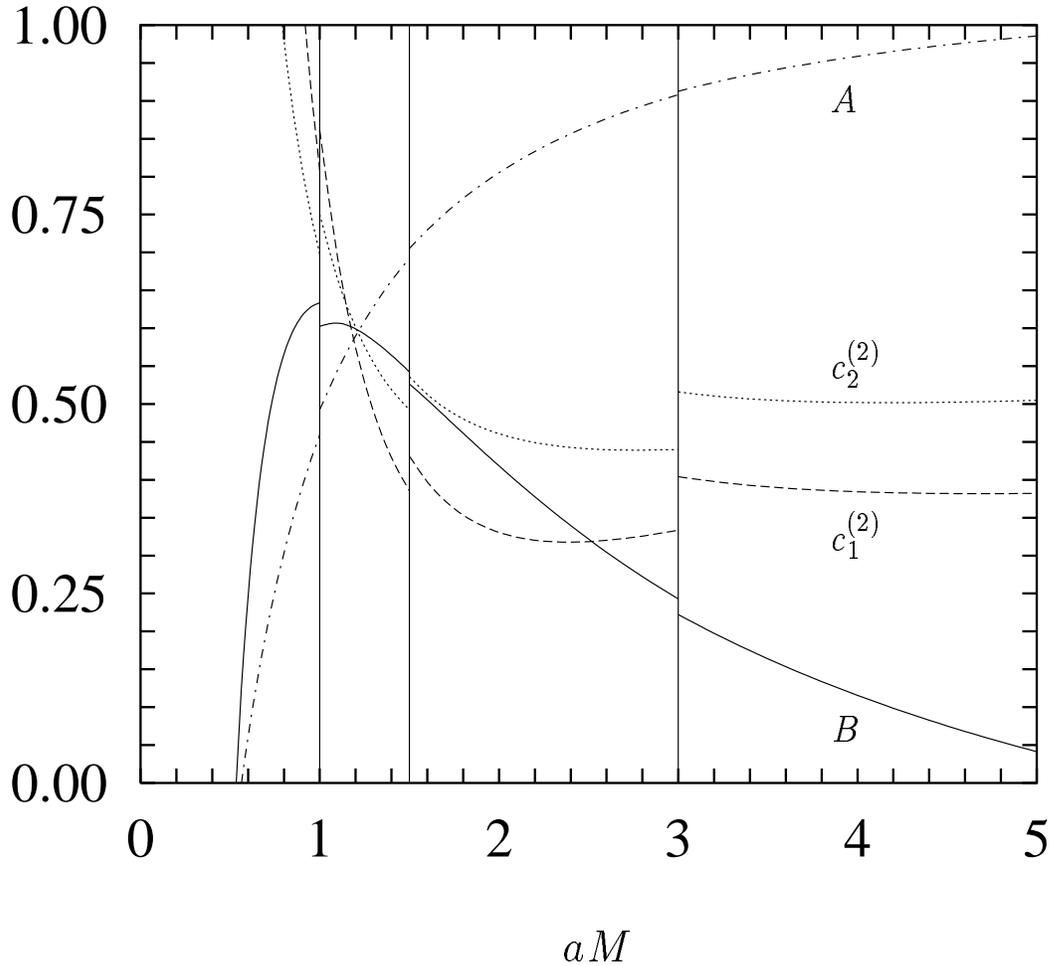}
\end{center}
\caption[figticoefs]
{The shift $A$ in the zero point of energy, the heavy-quark mass
renormalization parameter $B$, and the kinetic coupling coefficients
$c_1^{(2)}$ and $c_2^{(2)}$ {\em after} tadpole improvement
($u_0=1-\alpha_s\pi/3$) against the product of the bare heavy-quark
mass $M$ and the lattice spacing $a$ using $S_G^{(W)}$ and $\delta H^{(2)}$.
The dot-dashed curve is  $A$, the solid curve is $B$, the dashed curve
is $c_1^{(2)}$, and the dotted curve indicates $c_2^{(2)}$.  The stability
parameter $n$ assumes the same values as used in Fig.~\ref{figcoefs}.}
\label{figticoefs}
\end{figure}

\begin{figure}
\begin{center}
\leavevmode
\epsfbox[80 360 530 760]{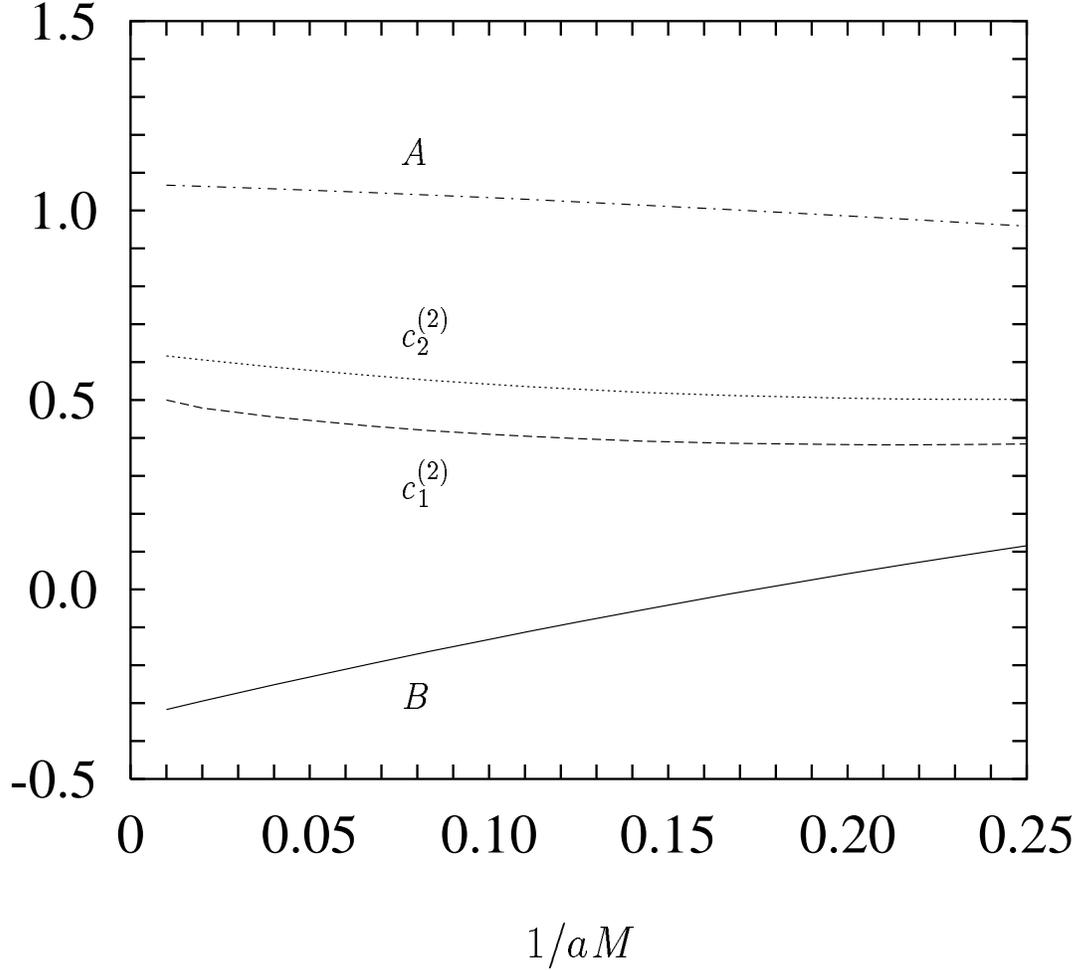}
\end{center}
\caption[figticoefs2]
{The shift $A$ in the zero point of energy, the heavy-quark mass
renormalization parameter $B$, and the kinetic coupling coefficients
$c_1^{(2)}$ and $c_2^{(2)}$ {\em after} tadpole improvement
($u_0=1-\alpha_s\pi/3$) against $1/aM$ using $S_G^{(W)}$ and $\delta H^{(2)}$.
The dot-dashed curve is  $A$, the solid curve is $B$, the dashed curve
is $c_1^{(2)}$, and the dotted curve indicates $c_2^{(2)}$.  The stability
parameter $n$ is set to unity.}
\label{figticoefs2}
\end{figure}

\begin{figure}
\begin{center}
\leavevmode
\epsfbox[80 360 530 760]{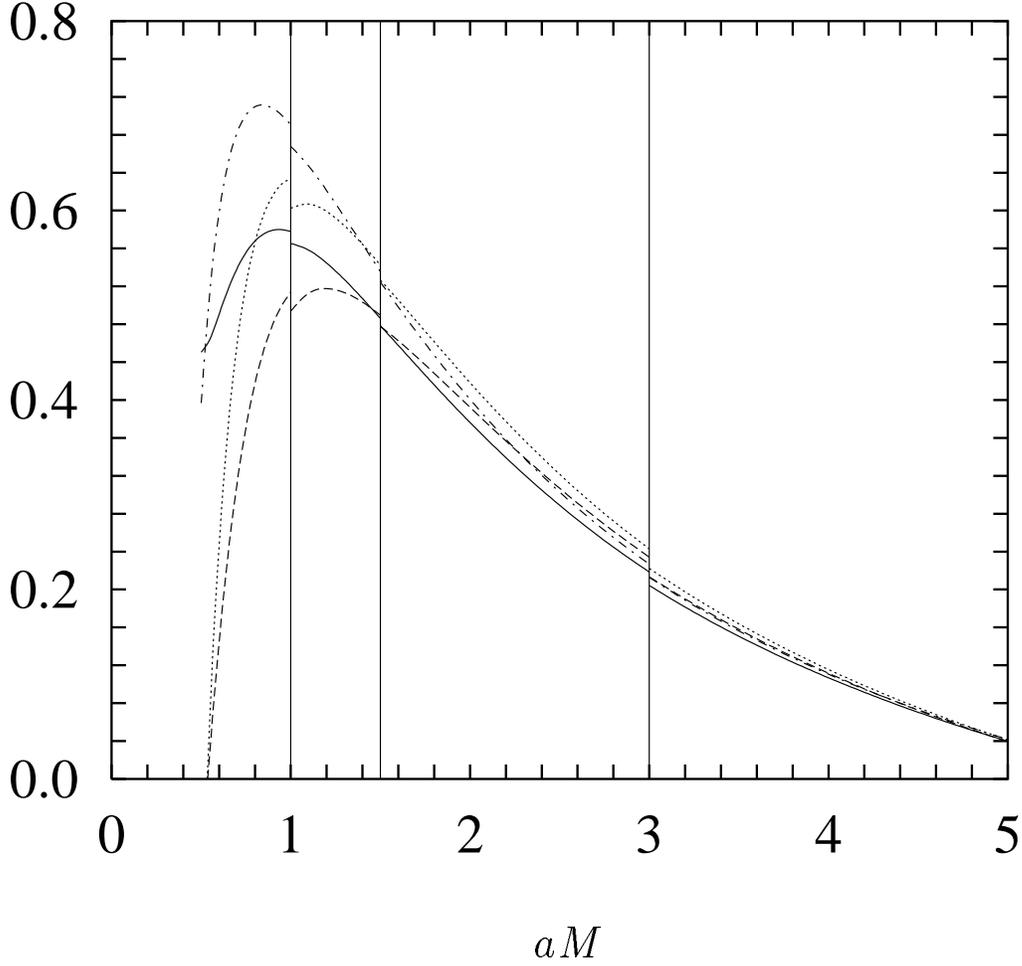}
\end{center}
\caption[figmass]
{The heavy-quark mass renormalization parameter $B$
against the product of the bare heavy-quark mass $M$ and the
lattice spacing $a$.  Results using the improved gluon action $S_G^{(I)}$
and the full set of NRQCD interactions $\delta H^{(4)}$ are shown
as a solid curve.  Results using $S_G^{(I)}$ and $\delta H^{(2)}$ are
shown as a dot-dashed curve.  The dashed curve indicates the results
obtained using the simple gluon action $S_G^{(W)}$ with $\delta H^{(4)}$,
while the dotted curve shows the results using $S_G^{(W)}$ with
$\delta H^{(2)}$.  The stability parameter $n$ assumes the same values as
used in Fig.~\ref{figcoefs}.}
\label{figmass}
\end{figure}

\begin{figure}
\begin{center}
\leavevmode
\epsfbox[80 360 530 760]{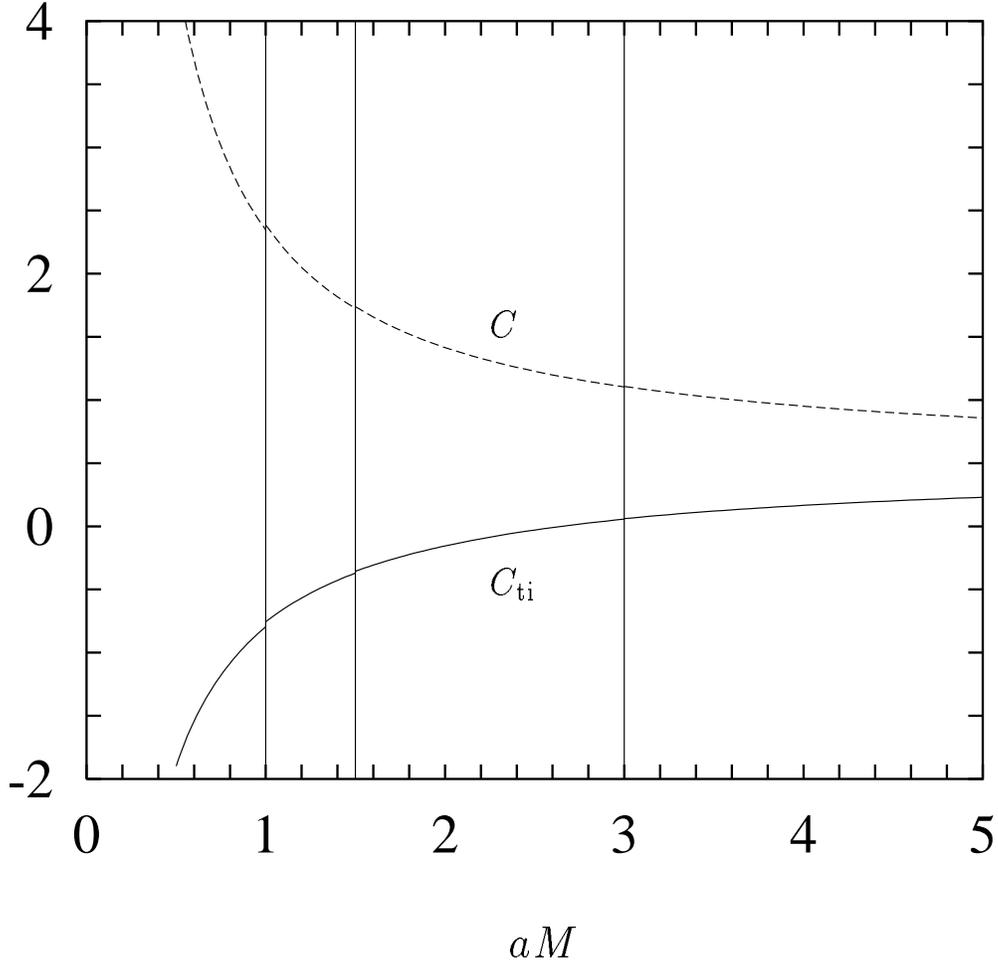}
\end{center}
\caption[figwavef]
{The heavy-quark wave function renormalization parameter $C$ against
the product of the bare heavy-quark mass $M$ and the lattice spacing $a$
using $S_G^{(W)}$ and $\delta H^{(2)}$.  The dashed curve shows $C$ before
tadpole improvement $(u_0=1)$ and the solid curve shows
$C$ after mean-field improvement $(u_0=1-\alpha_s\pi/3)$. The stability
parameter $n$ assumes the same values as used in Fig.~\ref{figcoefs}.}
\label{figwavef}
\end{figure}

The shift $A$ in the zero point of energy, the mass renormalization
$B$, and the kinetic coupling corrections $c_1^{(2)}$ and $c_2^{(2)}$
{\em before} tadpole improvement $(u_0^{(2)}=0)$ are shown in
Figs.~\ref{figcoefs} and \ref{figcoefs2} for the simple Wilson gluonic
action $S_G^{(W)}$ and the $O(v^2)$ version $\delta H^{(2)}$
of NRQCD \cite{otherres}.  Their values after tadpole improvement
are shown in Figs.~\ref{figticoefs} and \ref{figticoefs2}.
These quantities are gauge invariant and infrared finite.  In order
to demonstrate the sensitivity of the tadpole-improved quantities to the
form of the gluon action and the higher-order NRQCD interactions,
results for $B$ after tadpole improvement using both versions
$\delta H^{(2)}$ and $\delta H^{(4)}$ of NRQCD
and the simple Wilson $S_G^{(W)}$
and the $O(a^2)$-improved $S_G^{(I)}$ gluonic actions are shown in
Fig.~\ref{figmass}.  The wave function renormalization parameter $C$
before and after mean-field improvement for $S_G^{(W)}$ and $\delta H^{(2)}$
is presented in Fig.~\ref{figwavef}.

Before tadpole improvement, these parameters are all very large, especially
as $aM$ decreases.  Tadpole contributions dominate and contain power-law
divergences which grow as $aM$ becomes small.  Since high-momentum modes
are more strongly damped in the improved gluon propagator, the
ultraviolet-divergent tadpole terms, and hence the couplings and
renormalization parameters before tadpole improvement, are smaller
in the case of the improved gluon action.

Mean-field improvement greatly reduces the magnitudes of all of
these coefficients.  For $aM>5$, $A$ is almost independent of $aM$ after
tadpole improvement, with values near unity; as $aM$ decreases below $5$,
$A$ also decreases, eventually changing in sign.  After tadpole
improvement, $B$ has a small negative value at large $aM$; as $aM$
decreases, $B$ increases, becoming positive, reaches a maximum near $aM\sim 1$,
then rapidly decreases, eventually falling below zero again.
The kinetic coupling coefficients $c_1^{(2)}$ and $c_2^{(2)}$ vary
slowly for $aM\gtrsim 3/2$, then rise sharply as $aM$ decreases below
this.  $C$ also varies slowly, remaining a small correction, for
$aM\gtrsim 3/2$ after mean-field improvement.  As $aM$ decreases below
unity, however, $C$
begins to grow quickly in magnitude.  For large $aM$, these tadpole-improved
parameters are only slightly sensitive to the value of $n$, the spin-dependent
interactions, the cutoff improvement of the gluon action, and the $O(v^4)$
relativity corrections of NRQCD (see Fig.~\ref{figmass}).  As $aM$
decreases, however, sensitivity to these factors grows.
The kinetic couplings are generally more sensitive to these effects
than are the renormalization parameters $A$, $B$, and $C$.
The mass renormalization parameter $B$ is also very sensitive to the
leading kinetic interactions $V_1$ and $V_2$.

\subsection{Scale determinations}
\begin{figure}
\begin{center}
\leavevmode
\epsfbox[80 360 530 760]{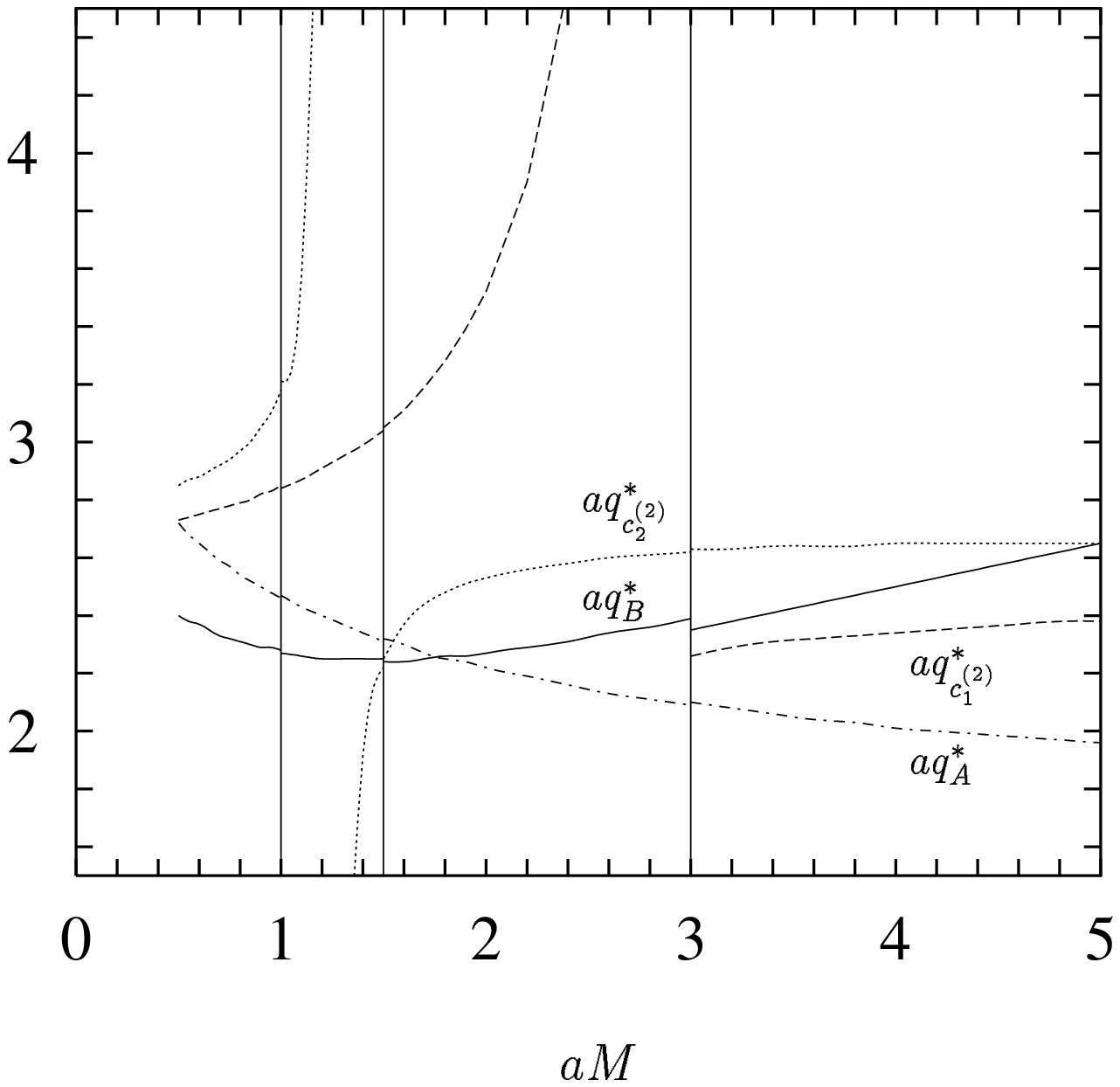}
\end{center}
\caption[figscales]
{The energy shift scale $q^\ast_A$, heavy-quark mass renormalization
scale $q^\ast_B$, and the kinetic coupling coefficient scales
$q^\ast_{c_1^{(2)}}$ and $q^\ast_{c_2^{(2)}}$ {\em before} tadpole
improvement ($u_0=1$) against the product of the bare heavy-quark mass
$M$ and the lattice spacing $a$ using $S_G^{(W)}$ and $\delta H^{(2)}$.
The dot-dashed curve is $aq^\ast_A$, the solid curve
shows $aq^\ast_B$, the dashed curve is $aq^\ast_{c_1^{(2)}}$,
and the dotted curve indicates $aq^\ast_{c_2^{(2)}}$.
The stability parameter $n$ assumes the same values as used
in Fig.~\ref{figcoefs}.}
\label{figscales}
\end{figure}

\begin{figure}
\begin{center}
\leavevmode
\epsfbox[80 360 530 760]{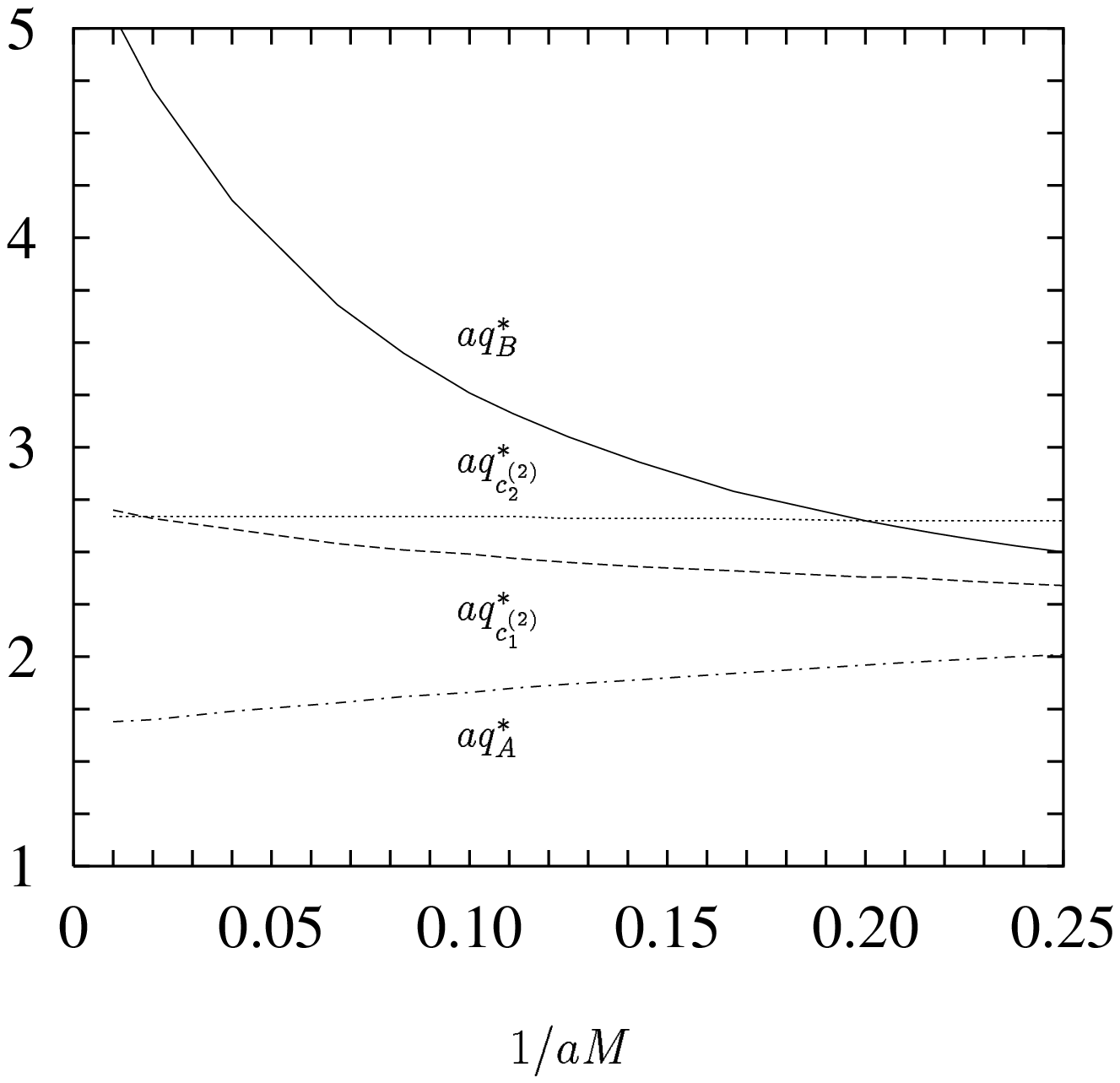}
\end{center}
\caption[figscales2]
{The energy shift scale $q^\ast_A$, heavy-quark mass renormalization
scale $q^\ast_B$, and the kinetic coupling coefficient scales
$q^\ast_{c_1^{(2)}}$ and $q^\ast_{c_2^{(2)}}$ {\em before} tadpole
improvement ($u_0=1$) against $1/aM$ using $S_G^{(W)}$ and $\delta H^{(2)}$.
The dot-dashed curve is $aq^\ast_A$, the solid curve
shows $aq^\ast_B$, the dashed curve is $aq^\ast_{c_1^{(2)}}$,
and the dotted curve indicates $aq^\ast_{c_2^{(2)}}$.
The stability parameter $n$ is unity.}
\label{figscales2}
\end{figure}

\begin{figure}
\begin{center}
\leavevmode
\epsfbox[80 360 530 760]{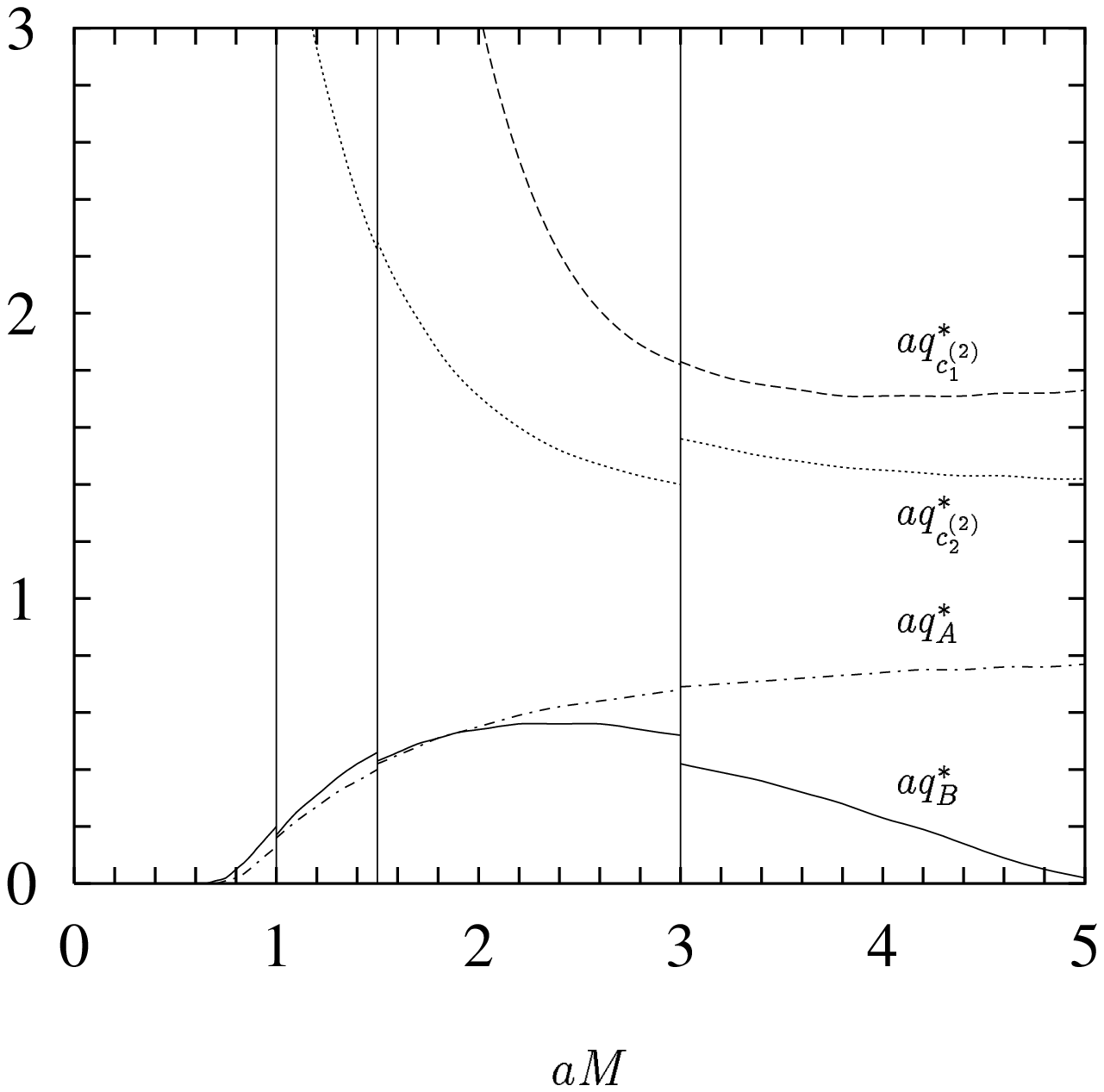}
\end{center}
\caption[figtiscales]
{The energy shift scale $q^\ast_A$, heavy-quark mass renormalization
scale $q^\ast_B$, and the kinetic coupling coefficient scales
$q^\ast_{c_1^{(2)}}$ and $q^\ast_{c_2^{(2)}}$ {\em after} tadpole
improvement ($u_0=1-\alpha_s\pi/3$) against the product of the bare
heavy-quark mass $M$ and the lattice spacing $a$ using $S_G^{(W)}$ and
$\delta H^{(2)}$.  The dot-dashed curve is $aq^\ast_A$, the solid curve
shows $aq^\ast_B$, the dashed curve is $aq^\ast_{c_1^{(2)}}$,
and the dotted curve indicates $aq^\ast_{c_2^{(2)}}$.
The stability parameter $n$ assumes the same values as used
in Fig.~\ref{figcoefs}.}
\label{figtiscales}
\end{figure}

\begin{figure}
\begin{center}
\leavevmode
\epsfbox[80 360 530 760]{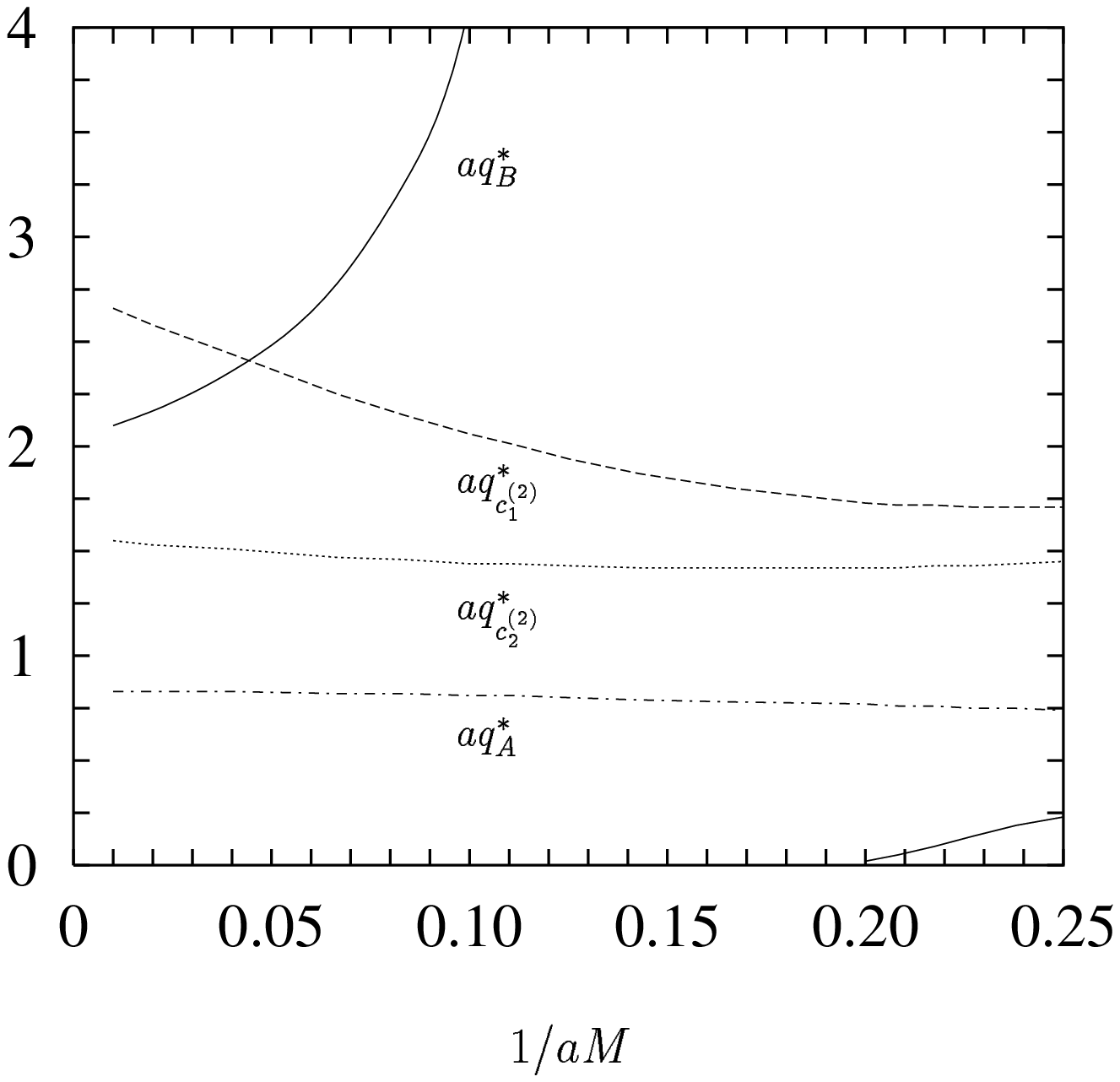}
\end{center}
\caption[figtiscales2]
{The energy shift scale $q^\ast_A$, heavy-quark mass renormalization
scale $q^\ast_B$, and the kinetic coupling coefficient scales
$q^\ast_{c_1^{(2)}}$ and $q^\ast_{c_2^{(2)}}$ {\em after} tadpole
improvement ($u_0=1-\alpha_s\pi/3$) against $1/aM$ using $S_G^{(W)}$ and
$\delta H^{(2)}$.  The dot-dashed curve is $aq^\ast_A$, the solid curve
shows $aq^\ast_B$, the dashed curve is $aq^\ast_{c_1^{(2)}}$,
and the dotted curve indicates $aq^\ast_{c_2^{(2)}}$.
The stability parameter $n$ is unity.}
\label{figtiscales2}
\end{figure}

The scales {\em before} tadpole improvement $(u_0^{(2)}=0)$ are shown
in Figs.~\ref{figscales} and \ref{figscales2} for the simple Wilson
gluonic action $S_G^{(W)}$ and the $O(v^2)$ version $\delta H^{(2)}$ of NRQCD
\cite{otherres}.  Their values {\em after} mean-field improvement are
shown in Figs.~\ref{figtiscales} and \ref{figtiscales2}.
The qualitative behavior of these scales are not affected by
the cutoff improvement of the gluon action and the $O(v^4)$
relativity corrections of NRQCD.  For large $aM$, the tadpole-improved
scales are especially insensitive to these factors.  As $aM$ decreases,
however, quantitative sensitivity grows, becoming rather pronounced
for $q^\ast_{c_1^{(2)}}$ and $q^\ast_{c_2^{(2)}}$.

Before tadpole improvement, the scales $q_A^\ast$ and $q_B^\ast$ are slowly
varying functions of $aM$ for $aM<5$, ranging between $2/a$ and $3/a$ in
value.  As $aM$ becomes large, $q_A^\ast$ falls slowly to a value just
below $2/a$, while $q_B^\ast$ rises to a value above $5/a$.
Since the unimproved parameters are dominated by highly-ultraviolet tadpole
contributions, scales in the range $2/a$ to $3/a$ are not surprising.
The greater damping of the high-momentum modes in the case of the improved
gluon action $S_G^{(I)}$ results in slightly lower scales in comparison
to those obtained using the Wilson action.

The scales $q_{c_1^{(2)}}^\ast$ and $q_{c_2^{(2)}}^\ast$ also vary
slowly between $2/a$ and $3/a$, except in the regions where the perturbative
coefficients $c_1^{(2)}$ and $c_2^{(2)}$ are nearly zero.  As these zeros
are approached from below, the scales tend asymptotically to infinity;
approached from above, the scales tend to zero.  This dramatic behavior
is an inevitable consequence of using Eq.~\ref{scaleform} to define the
scale and does not stem from any physical effect; whenever any perturbative
coefficient nears zero, similar pathological behavior will be observed in
the corresponding scale.  Such points at which the scales tend to zero on
one side and asymptotically to infinity on the other shall be referred to
here as {\em defects}.

For $aM>3$, the scale $q_A^\ast$ is almost independent of $aM$ with values
near $0.8/a$ after mean-field improvement, revealing a significant
reduction from the tadpole removal.  As $aM$ decreases below $3$, $q_A^\ast$
falls, becoming small as the zero in $A$ near $aM\sim 0.6$ is approached.
For $aM$ below this zero, $q_A^\ast$ assumes large defective values
which cannot be shown in Fig.~\ref{figtiscales}.  The mass renormalization
scale $q_B^\ast$ becomes small as the zero in $B$ near $aM\sim 0.5$ is
approached from above and the zero near $aM\sim 5$ is approached from below;
it becomes infinitely large as one approaches these zeros from the
opposite directions.  Between these defects, $q_B^\ast$ attains a maximum
value of only about $0.6/a$, and as $aM$ becomes large, $q_B^\ast$ tends to
a value near $2/a$; both of these facts indicate a substantial lowering of
$q_B^\ast$ from tadpole improvement.

For $aM>3$, the scale $q_{c_2^{(2)}}^\ast$ is significantly smaller
after mean-field improvement and varies little;
as $aM$ decreases below $3$, $q_{c_2^{(2)}}^\ast$ grows quickly,
eventually reaching values above $6/a$, indicating an enhancement from
tadpole removal.  Note that mean-field improvement has eliminated all
pathological values. $q_{c_1^{(2)}}^\ast$ displays behavior similar
to that of $q_{c_2^{(2)}}^\ast$, except that $q_{c_1^{(2)}}^\ast$ grows
to much greater values as $aM$ decreases, exceeding $15/a$ near
$aM\sim 1$ before leveling off.  These high values are a result of
large, nearly canceling contributions in the
loop integrals in Eq.~\ref{scaleform}.

\subsection{Size of the corrections}
In addition to the perturbative coefficients $A$, $B$, $c_1^{(2)}$,
$c_2^{(2)}$, and their associated scales, the QCD scale parameter
$a\Lambda_V$ must be known before numerical estimates of the first-order
energy shift $aE_0=A\ \alpha_V(q^\ast_A)$, the mass renormalization
$Z_m-1=B\ \alpha_V(q_B^\ast)$, and the radiative corrections $\delta c_1
=c_1^{(2)}\ \alpha_V(q^\ast_{c_1^{(2)}})$ and $\delta c_2
=c_2^{(2)}\ \alpha_V(q^\ast_{c_2^{(2)}})$ can be determined. The
most straightforward procedure for measuring $a\Lambda_V$
is to extract $\alpha_V(\mu)$ at some scale $\mu$ from simulations.
One possibility is to fit simulation measurements of the logarithm
of the mean plaquette to its perturbative expansion.  Doing this,
Lepage and Mackenzie \cite{lep92b} have extracted $\alpha_V(3.40/a)$
from Monte Carlo data for quenched QCD at several values of $\beta$,
where $\beta=6/g^2$.  At $\beta=6$, they find $\alpha_V(3.40/a)=0.152$,
corresponding to $a\Lambda_V=0.169$ in Eq.~\ref{twolooprun}; at
$\beta=6.4$, they obtain $\alpha_V(3.40/a)=0.130$, yielding
$a\Lambda_V=0.097$.

\begin{figure}
\begin{center}
\leavevmode
\epsfbox[80 360 530 760]{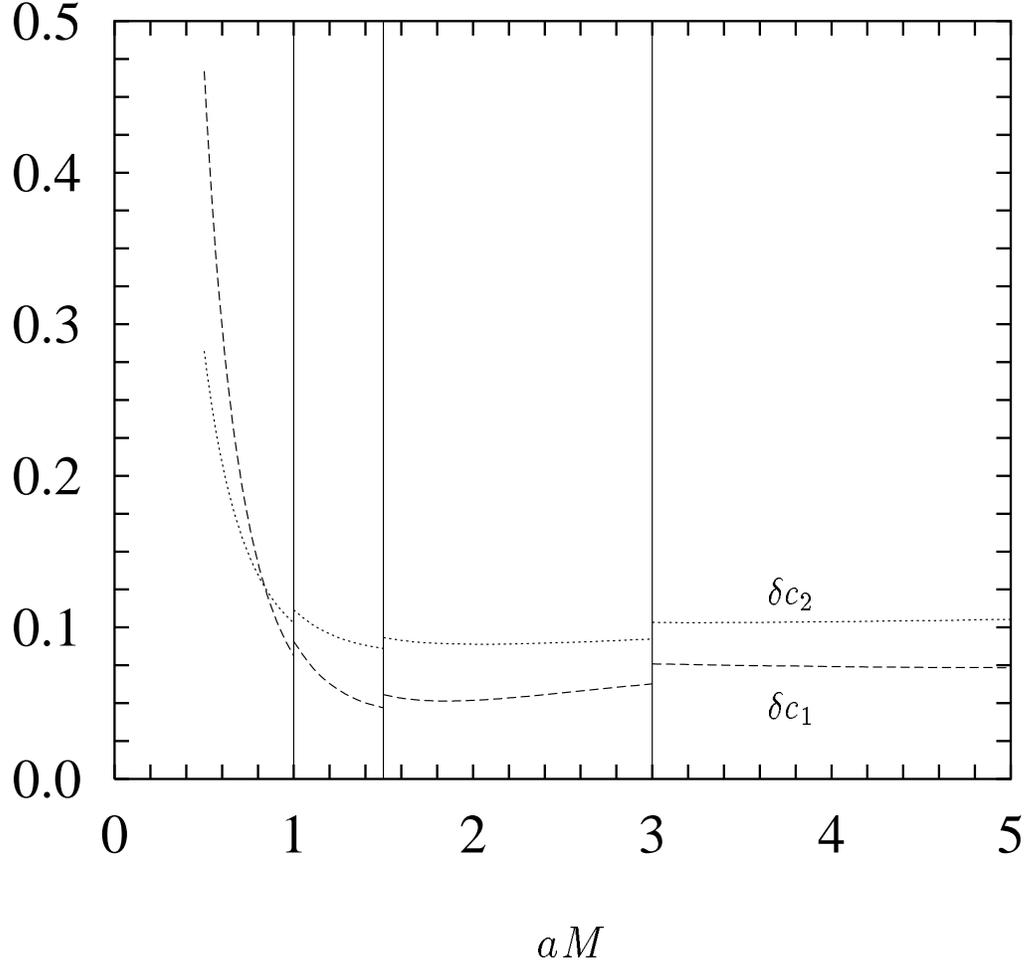}
\end{center}
\caption[figkincoefs]
{The first-order radiative corrections to the kinetic couplings
$c_1$ and $c_2$ against $aM$ using $S_G^{(W)}$ and $\delta H^{(2)}$.  The
dashed curve is $\delta c_1=c_1^{(2)}\ \alpha_V(q^\ast_{c_1^{(2)}})$
and the dotted curve indicates $\delta c_2=c_2^{(2)}\
\alpha_V(q^\ast_{c_2^{(2)}})$, using $a\Lambda_V=0.169$.
The stability parameter $n$ assumes the same values as used
in Fig.~\ref{figcoefs}.}
\label{figkincoefs}
\end{figure}

The leading-order radiative corrections $\delta c_1$ and $\delta c_2$
to the kinetic couplings are shown in Fig.~\ref{figkincoefs}
for $a\Lambda_V=0.169$ and using $\delta H^{(2)}$
and $S_G^{(W)}$.  For $aM>1$, $\delta c_1$ and $\delta c_2$ are
nearly independent of $aM$ and adjust $c_1$ and $c_2$ above their
tree-level strengths by about 10\%, a small amount.  For $aM<1$,
these corrections grow rapidly as $aM$ decreases, indicating the
insufficiency of tree-level values for $c_1$ and $c_2$ as $aM$ becomes
small.  These large corrections also suggest that the NRQCD approach
should be applied cautiously as $aM$ falls below unity.

The defects in $q^\ast_A$ and $q^\ast_B$ make determination of $aE_0$
and $Z_m-1$ problematical.  One cannot simply use $aE_0=A\ \alpha_V
(q^\ast_A)$ and $Z_m-1=B\ \alpha_V(q^\ast_B)$ near the defects
since $\alpha_V(q)$ is not reliably known for $q$ below and near
$\Lambda_V$.  Imposing minimum and maximum scales also does not work
since this produces large, unphysical discontinuities in $aE_0$ and
$Z_m-1$ on crossing the defects.  The $q^\ast$ prescription of
Eq.~\ref{scaleform} fails near the defects and it would seem that
one must resort to guessing the appropriate scale.  One suggestion
for doing this is to inspect the scale plots and somehow smooth
out the scales across the defects.  However chosen, denote this
guessed scale by $\bar q$ to distinguish it from the Lepage-Mackenzie
scale $q^\ast$.  Estimates of $aE_0$ and $Z_m-1$ can then be obtained
by replacing $\alpha_V(q^\ast)$ by $\alpha_V(\bar q)$.  A better
procedure, however, is to replace $\alpha_V(q^\ast)$ by its expansion
in terms of the coupling renormalized at $\bar q$, namely,
\begin{equation}
\bar\alpha_V(q^\ast;\bar q) = \alpha_V(\bar q)\left[1-\beta_0
 \ln\left(q^{\ast 2}/\bar q^2\right)\alpha_V(\bar q)\right].
\label{coupapprox}
\end{equation}
This incorporates information contained in $q^\ast$, yet yields
results which are continuous across the defects.

\begin{figure}
\begin{center}
\leavevmode
\epsfbox[80 360 530 760]{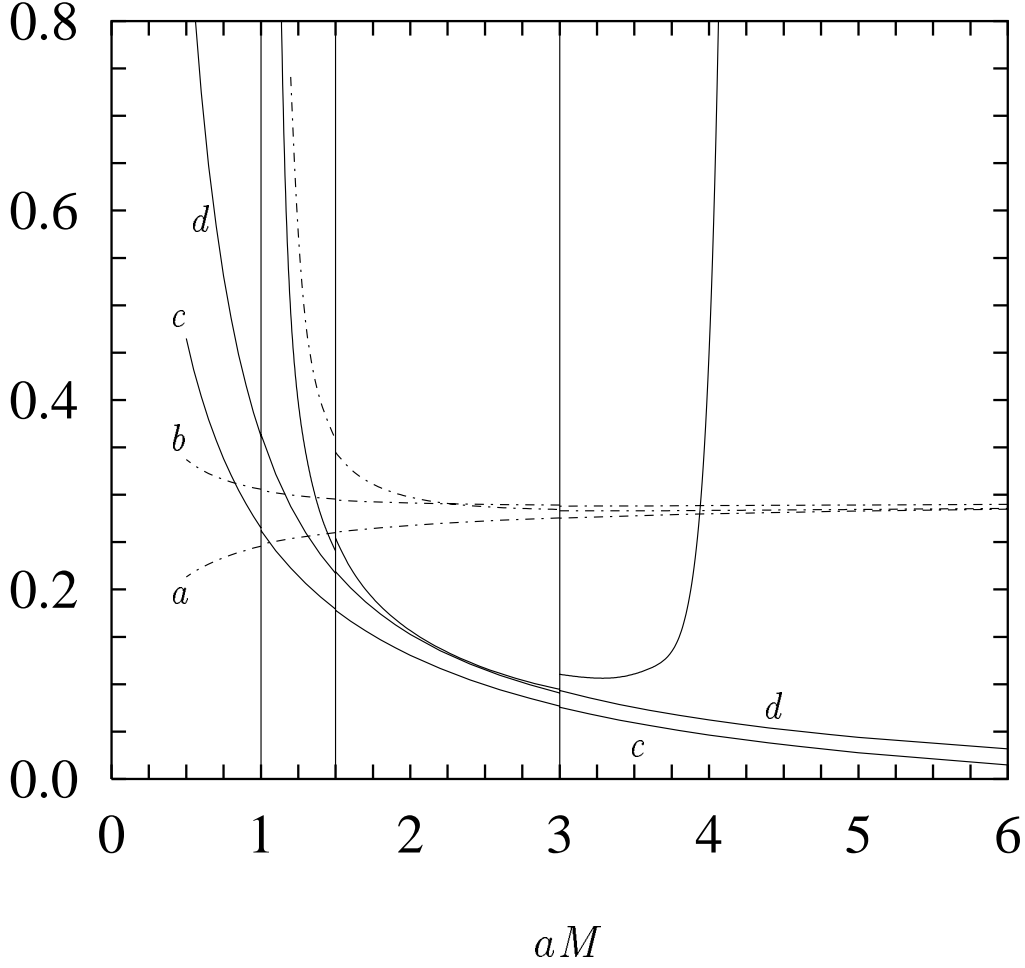}
\end{center}
\caption[figrenorms]
{Estimates of the energy shift $aE_0$ and mass renormalization $Z_m-1$
against $aM$ using $S_G^{(W)}$ and $\delta H^{(2)}$.  The dot-dashed curves
labeled $a$ and $b$ are energy shift estimates $A\ \bar\alpha_V(q^\ast_A;
\bar q_A)$ for $\bar q_A=0.8/a$ and $0.6/a$, respectively, using
$a\Lambda_V=0.169$ in $\alpha_V(\bar q_A)$;  the unlabeled dot-dashed
curve is $A\ \alpha_V(q_A^\ast)$.  The solid curves labeled
$c$ and $d$ show $B\ \bar\alpha_V(q^\ast_B;\bar q_B)$ for $\bar q_B=1.0/a$
and $0.6/a$, respectively; the unlabeled solid curve is
$B\ \alpha_V(q^\ast_B)$. Values for the stability parameter $n$ are the
same as in Fig.~\ref{figcoefs}.}
\label{figrenorms}
\end{figure}

A range of estimates for the energy shift
$aE_0$ and mass renormalization $Z_m-1$
using Eq.~\ref{coupapprox} are presented in Fig.~\ref{figrenorms}.
Energy shift estimates $A\ \bar\alpha_V(q^\ast_A;\bar q_A)$ are shown
choosing $\bar q_A=0.8/a$ and $0.6/a$ for all $aM$ values and
using $a\Lambda_V=0.169$ in $\alpha_V(\bar q_A)$.  Mass renormalization
estimates $B\ \bar\alpha_V(q^\ast_B;\bar q_B)$ are shown for the
choices $\bar q_B=1.0/a$ and $0.6/a$ for all $aM$.  These choices
for $\bar q_A$ and $\bar q_B$ are based on an examination of the
scales in Figs.~\ref{figtiscales} and \ref{figtiscales2}.
The energy shift $aE_0$ is remarkably independent of $aM$, in contrast
to the perturbative coefficient $A$ which falls to below zero as $aM$
decreases.  Sensitivity to the choice of $\bar q_A$ grows as $aM$
decreases.  The mass renormalization is very small for $aM>3$ and is
only mildly sensitive to the choice of $\bar q_B$; $Z_m-1$ becomes zero
near where the perturbative coefficient $B$ vanishes.  As $aM$
decreases below this, however, $Z_m-1$ grows quickly and continues
to grow even after $B$ has fallen to below zero.  As $aM$ becomes small,
estimates of the mass renormalization become very large and the
sensitivity of these estimates to $\bar q_B$ becomes pronounced.
Both of these facts and the sensitivity of the energy shift to $\bar q_A$
for small $aM$ underscore the need to compute the two-loop
corrections to $Z_m-1$ and $aE_0$.

\section{Conclusion}
\label{sec:concl}

The heavy-quark propagator in NRQCD was calculated to $O(\alpha_s)$
in tadpole-improved perturbation theory.  The mass renormalization,
shift in the zero point of energy, and the radiative corrections to
the two kinetic couplings $c_1$ and $c_2$ were determined by matching
the on-mass-shell dispersion relation for the heavy quark obtained
from the NRQCD quark propagator to that of continuum QCD to $O(v^4)$.
In addition, the scales at which to evaluate the running coupling for
these quantities were obtained for the first time using the
Lepage-Mackenzie prescription.  Defects in these scales were
found.  The wave function renormalization was extracted from the
on-shell residue of the perturbative propagator.  Results were obtained
using two different gluon actions $S_G^{(W)}$ and $S_G^{(I)}$ and two
versions of NRQCD, $\delta H^{(2)}$ and $\delta H^{(4)}$, differing
only in the order of the relativity and finite-lattice-spacing
corrections retained.

Using a typical value for $a\Lambda_{\rm QCD}$ from recent simulation
measurements of $\alpha_s$, numerical estimates of the energy shift,
mass renormalization, and the radiative corrections to $c_1$ and $c_2$
were also obtained.  The radiative corrections to the kinetic couplings
were found to be small for $aM>1$, being roughly 10\% of
the tree-level coupling strengths, and to grow quickly
as $aM$ decreased below unity.  Problems in reliably setting the scale
complicated the determinations of the energy shift and mass renormalization
for some values of $aM$.  A simple modification of the running coupling
was necessary to obtain a range of estimates for these parameters near
these $aM$ values.  The mass renormalization was found to be small for
$aM>3$ and to become large as $aM$ decreased.  Large uncertainties in
determining the energy shift and mass renormalization for small $aM$
demonstrated the need to compute the two-loop corrections to these
quantities.

Future work includes the evaluation of the other NRQCD couplings
and the two-loop contributions to the mass renormalization and the energy
shift.  Computation of the couplings $c_3-c_7$ has already begun.

\acknowledgments
I would like to thank G.~P.~Lepage, C.~Davies, J.~Sloan,
J.~Shigemitsu, K.~Hornbostel and A.~Langnau for many useful
conversations.  This work was supported by the Natural Sciences
and Engineering Research Council of Canada,
the U.~S.~Department of Energy, Contract No.~DE-AC03-76SF00515,
and the UK Science and Engineering Research Council, grant
GR/J 21347.
\epsfverbosetrue

\begin{references}
\bibitem{change}
   This evolution equation differs slightly from that previously
   used in Refs.~\cite{lep92} and \cite{mor93}.  The NRQCD action
   was modified to simplify both perturbative and Monte Carlo
   computations.
\bibitem{mor93diff}
   Because the new NRQCD action is simpler than that studied
   previously, the procedure used here to evaluate the moments
   $\Omega_m^{(l)}$ differs slightly from that described in
   Ref.~\cite{mor93}.
\bibitem{otherres}
   Results for both $\delta H^{(2)}$ and $\delta H^{(4)}$
   using the simple Wilson $S_G^{(W)}$ and the
   $O(a^2)$-improved $S_G^{(I)}$ gluonic actions are available
   in tabular form from the author.
\end{references}
\end{document}